# Giant oscillatory tunnel magnetoresistance in CoFe/MgO/CoFe(001) junctions

Thomas Scheike,[1] Zhenchao Wen,[1] Hiroaki Sukegawa,[1,*] and Seiji Mitani[1]

[1]*National Institute for Materials Science (NIMS), Tsukuba 305-0047, Japan*

## Abstract

The tunnel magnetoresistance (TMR) effect observed in magnetic tunnel junctions (MTJs) is commonly used in many spintronic applications because the effect can easily convert from local magnetic states to electric signals in a wide range of device resistances. In this study, we demonstrated TMR ratios of up to 631% at room temperature (RT), which is two or more times larger than those used currently for magnetoresistive random access memory (MRAM) devices, using CoFe/MgO/CoFe(001) epitaxial MTJs. The TMR ratio increased up to 1143% at 10 K, which corresponds to an effective tunneling spin polarization of 0.923. The observed large TMR ratios resulted from the fine-tuning of atomic-scale structures of the MTJs, such as crystallographic orientations and MgO interface oxidation, in which the well-known $\Delta_1$ coherent tunneling mechanism for the giant TMR effect is expected to be pronounced. However, behavior that is not covered by the standard coherent tunneling theory was unexpectedly manifested; i.e., (i) TMR saturation at a thick MgO barrier region and (ii) enhanced TMR oscillation with a 0.32 nm period in MgO thickness. Particularly, the TMR oscillatory behavior dominates the transport in a wide range of MgO thicknesses; the peak-to-valley difference of the TMR oscillation exceeded 140% at RT, attributable to the appearance of large oscillatory components in resistance area product ($RA$). Further, we found that the oscillatory behaviors of the TMR ratio and $RA$ survive, even under a ±1 V bias voltage application, indicating the robustness of the oscillation. Our demonstration of the giant RT-TMR ratio will be an essential step for establishing spintronic architectures, such as large-capacity MRAMs and spintronic artificial neural networks. More essentially, the present observations can trigger us to revisit the true TMR mechanism in crystalline MTJs.

Corresponding author: SUKEGAWA.Hiroaki@nims.go.jp

# I. INTRODUCTION

The tunnel magnetoresistance (TMR) effect, which is based on spin-dependent tunneling between ferromagnets (FMs), has been studied in applied physics for over 40 years.[1] The TMR effect has been extensively studied from both theoretical and experimental perspectives and has been successfully implemented in practical device applications.[2] However, recent experimental results are incompatible with common theoretical models, such as TMR ratios that are much smaller than predicted and an oscillatory behavior with the barrier layer thickness that cannot be reasonably reproduced by calculations.

The TMR effect is the core operation principle of numerous spintronic devices, such as hard disk drives (HDDs) and nonvolatile magnetoresistive random access memories (MRAMs). This effect is observed in magnetic tunnel junctions (MTJs) comprising an FM/ultrathin insulator (barrier)/FM core structure. The tunneling resistance of MTJs can easily vary without any atomic displacement only by controlling the relative direction of magnetizations of two FM layers using, e.g., an external magnetic field, spin-transfer torque,[3] and spin-orbit torque effects.[4] This feature allows (i) the detection of FM layers' magnetic states and external magnetic fields, (ii) the nonvolatile storage of digital data by restricting magnetized directions [i.e., parallel (P) and antiparallel (AP) states], and (iii) the infinite rewriting of data. Moreover, the MTJ resistance can be widely tuned by changing the barrier thickness, and the TMR effect can survive, even in nanometer-scale MTJs.[5] Therefore, MTJs are used in various applications, such as read-heads of HDDs, memory cells of MRAMs, and highly-sensitive magnetic sensors.

The magnitude of the resistance change by the TMR effect is an essential property of a MTJ, as it determines the electric output as an electronic device element. The ratio of the change between the P and AP states of FMs is the TMR ratio. The first TMR effect was demonstrated by Julliere in 1975;[1] a 14% TMR ratio was observed using a Fe/Ge/Co trilayer at low temperature and zero-bias voltage, and the basic concept of a two-current model for TMR was proposed (Julliere model). The model expects resistance change, i.e., TMR ratio, by assuming transport between FM layers with spin-conservation for each spin-band. Using an effective tunneling spin polarization $P_{\text{eff}}$, which is the spin polarization of a current flow through a barrier, the TMR ratio (%) can be expressed as $100 \times 2P_{\text{eff}}^2 / (1 - P_{\text{eff}}^2) = 100 \times (R_{\text{AP}} - R_{\text{P}}) / R_{\text{P}}$, where $R_{\text{AP}}$ ($R_{\text{P}}$) is the tunneling resistance of the AP (P) state. This indicates that a perfect spin polarization, i.e., $|P_{\text{eff}}| = 1$, results in an infinite TMR ratio. However, practical TMR ratios fall short of the theoretical infinite value; thus, $|P_{\text{eff}}|$ in actual MTJs is always much smaller than 1 due to many reasons. In the history of MTJs, TMR ratio keeps increasing.[6,7] After reporting over 10% TMR ratio at room temperature (RT) using MTJs with an amorphous $AlO_x$ barrier in 1995,[8,9] the development of spintronic devices, including HDD read-heads and MRAMs, significantly accelerated. The RT-TMR ratios significantly increased to 180%–220% through the crystalline MgO(001) barrier developed in 2004[10,11] after theoretical predictions on giant TMR ratios in 2001, which are based on a $\Delta_1$ band preferential coherent tunneling mechanism in a Fe/MgO/Fe(001) structure.[12,13] At present, almost all applications use MgO barrier-based MTJs due to the large TMR ratio.[6] The largest experimental TMR ratio at RT is 604%, which was reported in 2008 by Ikeda *et al.*[14] using a pseudo-spin-valve CoFeB/MgO/CoFeB MTJ. However, the progress in the RT-TMR ratio has stagnated over the last decade. Notably, practical MTJ-based devices can use only 100%–200% TMR ratios at best to simultaneously satisfy several requirements for their operation, such as circuit impedance matching, high-speed operation, good thermal stability, low power consumption for operation, and low $1/f$ noise at a low frequency range.[15–22]

A huge drawback of MTJs as an integrated circuit element is a small current on/off ratio (i.e., small TMR ratio). If much larger RT-TMR ratios are achieved, application ranges of TMR-based devices will significantly expand to very high-density MRAMs based on a three-dimensional architecture, nonvolatile magnetic logics,[23] brain-morphic devices,[24] etc. Some authors recently investigated simple Fe/MgO/Fe(001) epitaxial MTJs to clarify the large TMR ratio gap between experimental [180%–220% at RT, 250%–370% at low temperature (LT)][10,25,26] and theorical values (>1000%).[12,27–29] Significant improvement in the TMR ratio of up to 417% and 914% at RT and 3 K, respectively, in Fe/MgO/Fe MTJs by fine-tuning the crystallinity near the MgO barrier interfaces was demonstrated.[30] As such remarkable progress can be seen even in the simple Fe/MgO/Fe structure, it may be possible to significantly improve the RT-TMR ratio by effectively improving $P_{eff}$ by designing MTJ materials, such as barrier and FM layers, and controlling local barrier interface structures, such as the improvement of crystallinity and suppression of atomic defect formation.

Although recent studies on Fe/MgO/Fe have reported smaller TMR gap, other problems become more apparent. First, the TMR ratio is still small in experiments. In addition, the TMR ratio shows saturation behavior when the barrier thickness exceeds around 1–2 nm.[10,31–33] Based on the coherent tunneling theories, typical calculated TMR ratios are several thousand % for MTJs with 1–2-nm thick MgO barriers;[12,13] the ratios monotonically increase with the barrier thickness and reach ~10,000%–30,000% (on/off ratio: ~100–300) at a few nm.[12,29,34,35] Second, the TMR ratio frequently shows oscillatory behavior in experiments with barrier thickness in epitaxial MTJs. The oscillation period of ~0.3 nm is almost independent of the electrodes' FM materials, e.g., Fe,[10,30,36] $Co_2MnSi$ (CMS),[31] and $Co_2Cr_{0.4}Fe_{0.6}Al$ (CCFA).[32,33] A recent Fe/MgO/Fe MTJ showed an ~80% peak-to-valley (PV) difference in the RT-TMR ratio,[30] indicating the dominance of the transport properties by the oscillation effect. Further, a MTJ with a spinel-based barrier $Mg_4Al-O_x$, i.e., Fe/$Mg_4Al-O_x$/Fe(001), showed a similar ~0.3 nm period oscillation with an enhanced PV difference of ~125% at RT.[37] The physical origin of the oscillation is still under debate, although many theoretical studies have been conducted.[13,29,38,39] Therefore, the fundamental understanding of this oscillation behavior using devices with a very large TMR ratio is an important experimental task to clarify its origin. Investigations on these disparities between actual MTJs and theoretical calculations should also be critical for understanding the prerequisites for realizing huge RT-TMR ratios experimentally, such as >1000%.

In this study, we investigated TMR properties and oscillation behavior in detail to make the discrepancy between experiments and theories clear. First, we reported large TMR ratios using CoFe/MgO/CoFe(001) epitaxial MTJs by improving nanostructures of a MTJ stack structure. The RT-TMR ratio reached a maximum of 631%, which is the largest reported RT value in MTJs so far. The significant TMR enhancement originated from the introduction of sub-nm to a few nm CoFe layers at both the top- and bottom-side of the barrier interface of the Fe/MgO/Fe(001) framework structure. Second, we demonstrated a significantly large TMR oscillation with a maximum PV difference of 141% in the CoFe/MgO/CoFe(001) MTJs at RT. We found that the oscillation behavior was maintained, even under high bias voltages (>1 V), indicating that the origin of the oscillation is not explained only by the common $\Delta_1$ preferential coherent tunneling mechanism. Our demonstrations of the large RT-TMR ratio and its oscillatory behavior are crucial steps for developing novel spintronic applications in the future.

The rest of this article is organized as follows. In Sec. II, we show the experimental procedures for obtaining CoFe/MgO/CoFe(001) MTJs. The TMR properties at RT and LT and their bias voltage dependences are described in Sec. III. To obtain a full picture of the TMR oscillation behavior, we evaluated the MgO thickness dependences of $R_{AP}$, $R_P$, and the TMR ratio in a wide range of bias voltages, which is described in Sec. IV. The disparities between our experimental data and the theories of the $\Delta_1$ preferential coherent tunneling will be discussed in Sec. V.

## II. EXPERIMENTAL

The MTJ multilayer stack design was based on the previous Fe/MgO/Fe(001) structure with a 417% RT-TMR ratio.[30] The stacks were deposited using ultrahigh-vacuum multichamber deposition apparatus (base pressure: $4 \times 10^{-7}$ Pa) with direct current (DC)/radiofrequency magnetron sputtering and electron beam (EB) evaporation guns (ULVAC, Inc.). Typical stack structures were MgO(001) single crystal substrate//Cr (60)/Fe (50)/Co$_{50}$Fe$_{50}$ (CoFe) ($d_{bot\text{-}CoFe}$)/Mg ($d_{Mg}$)/wedge-shaped MgO ($d_{MgO}$ = 1.0 − 3.0)/natural oxidation/CoFe ($d_{top\text{-}CoFe}$)/Fe (5)/Ir$_{20}$Mn$_{80}$ (10)/Ru (20) [values in parentheses in nm, see Fig. 1 (a)]. The MgO barrier was deposited using the EB evaporation of a sintered MgO pellet and a linear shutter to create the wedge shape for the *x*-direction of a $20 \times 10$ mm$^2$ substrate [see the photographs of a MTJ wafer in Fig. 1 (b)]. After depositing the Cr-buffer, bottom-CoFe, MgO barrier, and top-Fe layer, *in situ* postannealing was performed at 600°C, 300°C, 250°C, and 400°C, respectively. More details on the film preparation are described in Ref.[30]. The single crystal cubic structures of bottom- and top-Fe electrodes and the MgO barrier were confirmed using reflection high energy electron diffraction (RHEED) observation [Fig. 1 (c)]. The multilayer wafers were *ex situ* annealed in a 0.2-T magnetic field along the MgO[110] || Fe[100] || CoFe[100] direction to induce exchange bias for the top-CoFe/Fe bilayer. The TMR ratios and resistance area products (*RA*s) of the wafers were evaluated using the current in-plane tunneling (CIPT) method (Capres A/S, CIPTech-SPM200 prober).[40] From Fig. 1 (d), the multilayer was patterned into MTJs with an ellipsoidal pillar shape ($10 \times 5$ μm$^2$ with the long axis along MgO[110]) using microfabrication techniques, e.g., photolithography, Ar ion etching, and magnetron sputtering for SiO$_2$ insulator and Au electrodes. The magnetotransport properties of the fabricated MTJs were characterized by a DC 4-probe method using a sourcemeter (Keihley, 2400) and nanovoltmeter (Keithley, 2182A) at RT and physical property measurement system (Quantum Design, Dynacool) for LT measurements. The TMR ratio was defined as $(R_{AP} - R_P) / R_P \times 100\%$, where $R_P$ ($R_{AP}$) is the resistance in the P (AP) magnetization state. At a positive bias voltage, electrons tunnel from the top- to the bottom electrode.

**III. RESULTS**

**A. TMR properties**

In a previous report of epitaxial Fe/MgO/Fe(001) MTJs, we demonstrated that MgO interface modification by introducing ultrathin layer insertion, postannealing, and postoxidation resulted in the achievement of symmetric current–voltage (*I–V*) characteristics, significantly improving TMR. In addition, a CoFe insertion at the MgO bottom interface enhanced the RT-TMR ratio, with 417% of no-insertion and 497% of a CoFe 2.24-nm insertion [= 16 monolayers (ML), where 1 ML of CoFe(001) = 0.143 nm]. Thus, we started with the framework structure of an underlayer/Fe (50 nm)/Mg (0.5 nm)/MgO/Fe (5 nm)/upper-layer used in Ref.[30] and modified the MTJ by tuning the MgO interface structures.

First, we investigated the bottom-CoFe insertion effect using the stack structure shown in the inset of Fig. 2 (a), i.e., Fe/CoFe ($d_{\text{bot-CoFe}}$)/Mg/MgO/Fe. In Fig. 2 (a), the 4-probe TMR ratio measured at RT with a bias voltage less than 10 mV as a function of $d_{\text{bot-CoFe}}$ in ML (bottom axis) and nm (top axis) of patterned MTJ pillars is shown. Here, each TMR ratio is the maximum value for each wafer. It is seen that the TMR ratio increases with $d_{\text{bot-CoFe}}$, and it saturates at around 16 ML. The maximum

TMR ratio was 504% at $d_{\text{bot-CoFe}}$ = 24 ML.

Next, the MgO interfaces were further tuned by controlling the CoFe layer thickness at the top-side ($d_{\text{top-CoFe}}$) and the Mg layer thickness at the bottom-side ($d_{\text{Mg}}$). Here, $d_{\text{top-CoFe}}$ = 16 ML was used as schematically shown in the inset of Fig. 2 (b), i.e., CoFe (16 ML)/Mg ($d_{\text{Mg}}$)/MgO/CoFe ($d_{\text{top-CoFe}}$). Figure 2 (b) shows the maximum RT-TMR ratio measured as a function of $d_{\text{top-CoFe}}$. When $d_{\text{top-CoFe}}$ was increased with maintaining $d_{\text{Mg}}$ = 0.5 nm (green line), the maximum TMR ratio of 551% was observed at $d_{\text{top-CoFe}}$ = 4 ML. MTJs with $d_{\text{top-CoFe}}$ higher than 8 ML showed smaller values, indicating that the CoFe insertion effect on a TMR ratio for the top interface differs from that for the bottom one. This behavior is attributable to the difference in their growth mode, which is sensitive to the state of the interfaces, e.g., interfacial roughness, degree of (001)-orientation, and interfacial strain. Figure 2 (b) also shows the relationship between the maximum TMR ratio and $d_{\text{Mg}}$. The TMR ratio further increased by optimizing $d_{\text{Mg}}$; namely, the maximum value of 631% at RT was observed at $d_{\text{top-CoFe}}$ = 4 ML and $d_{\text{Mg}}$ = 0.6 nm [black open squares in Fig. 2 (b)]. This TMR ratio is larger than the RT-TMR ratio of 604% reported by Ikeda *et al.*[14] using a CoFeB/MgO/CoFeB pseudo-spin-valve MTJ. Because the Mg insertion mainly works as a protection for the bottom electrode interface from oxidation in the deposition of the MgO barrier and postannealing process in our MTJs, $d_{\text{Mg}}$ = 0.6 nm is probably the optimum thickness that realizes favorable interface bonding states between CoFe and oxygen atoms at the MgO surface. A lower and higher $d_{\text{Mg}}$ reduced the TMR ratio due to over- and under-oxidation state at the bottom MgO interface, respectively. Notably, for different electrode materials, the optimum

Mg insertion thickness may change, e.g., $d_{Mg}$ = 0.5 nm for Fe electrode.[30]

The resistance (left axis) and TMR ratio (right axis) as a function of magnetic field ($H$) at RT for the MTJ showing the maximum TMR ratio of 631% ($d_{bot\text{-}CoFe}$ = 16 ML, $d_{top\text{-}CoFe}$ = 4 ML, $d_{Mg}$ = 0.6 nm, and $d_{MgO}$ = 1.86 nm) are shown in Fig. 2 (c). The curve shows a typical exchange-biased hysteresis loop with stable P and AP states at RT. The MTJ has a junction resistance = 70 Ω ($RA$ = 2.74 kΩ·μm²) for the P state, which increases to 7.31 times (= 514 Ω) when it switches to the AP state. To exclude any possible measurement or microfabrication errors, we also evaluated zero-bias TMR ratios and $RA$ values through CIPT measurements of the *unpatterned* wafer with $d_{bot\text{-}CoFe}$ = 16 ML, $d_{top\text{-}CoFe}$ = 4 ML, and $d_{Mg}$ = 0.6 nm. Figures 2 (d) and (e) show the CIPT results near the wafer position of $d_{MgO}$ = 1.9 nm: (d) the sheet resistance in the P state ($R_\square^{low}$) and (e) the current-in-plane (CIP) TMR ratio (MR$_{CIP}$) versus the mean probe pitch.[40] The fits (circles) by the theoretical equations match measured data (cross marks) well. We obtained reasonable values of TMR ratio = 617% and $RA$ = 3.4 kΩ·μm² by the fit. Therefore, RT-TMR ratios exceeding 600% were confirmed by both an unpatterned wafer (CIPT) and patterned MTJ pillars (DC 4-probe method). Notably, the other fit parameters, sheet resistances for the top electrode ($R_t$) and the bottom electrode ($R_b$), agree with the values of previous Fe/MgO/Fe having almost the same electrode configurations,[30] also supporting reasonable fits by CIPT for our MTJ stack. Additional CIPT results of MTJ wafers are shown in Supplementary Figs. S1–S3.

Figure 3 shows the temperature dependences of (a) the TMR ratio and (b) $R_P$ and $R_{AP}$ (bias voltage < 10 mV). The corresponding conductance ratio [= ($g_P − g_{AP}$) / $g_P$ = ($R_{AP} − R_P$) / $R_{AP}$, where $g_{P(AP)} \equiv 1 / R_{P(AP)}$ is the DC conductance in the P (AP) state] is also plotted on the right axis. The TMR and conductance ratios monotonically increase with a decrease in temperature. The TMR ratio reached a maximum value of 1143% at 10 K, which is much larger than the value of the previous Fe/MgO/Fe (914%)[30] and almost the same as the value of the CoFeB/MgO/CoFeB by Ikeda et al. at LT (1144%).[14]

The inset of Fig. 3 shows the corresponding TMR-$H$ loop at 10 K. Below 10 K, the TMR ratio reduces slightly because of an imperfect AP state, suggesting that the observed value is underestimated at the LT limit. Using the Julliere model [1] and TMR ratio = $100 \times 2P_{\text{eff}}^2/(1 - P_{\text{eff}}^2)$, $P_{\text{eff}}$ at RT (LT) was calculated to be 0.871 (0.923) by assuming both interfaces had the same $P_{\text{eff}}$. The ratio of $P_{\text{eff}}$ at RT to LT [$P_{\text{eff}}(RT)/P_{\text{eff}}(LT)$] was 0.94, which was higher than 0.91 of the previous Fe/MgO/Fe.[30] The weaker temperature dependence is attributable to the increased interface Curie temperature by Co doping into Fe,[41,42] which may effectively improve the interlayer exchange stiffness constant.[43] Recent theoretical work considering an intra-atomic $s$–$d$ exchange interaction also predicted monotonic temperature dependence improvement in CoFe/MgO/CoFe(001) with an increase in Co composition.[44] Such an improvement in the temperature dependence was reported in a bcc-Co/MgO/bcc-Co(001) MTJ [$P_{\text{eff}}(RT)/P_{\text{eff}}(LT) = 0.97$].[45] Therefore, introducing CoFe at both MgO interfaces effectively suppressed the temperature dependence of $P_{\text{eff}}$ in addition to the high $P_{\text{eff}}$ at LT, yielding a giant RT-TMR ratio in the present MTJ.

The temperature dependence of the TMR ratio follows that of $R_{AP}$ rather than $R_P$, which is commonly observed in various MTJs,[25,46,47] due to strong temperature dependence in $R_{AP}$ in contrast to very weak dependence in $R_P$. Interestingly, the temperature dependence of $R_P$ shows a complicated behavior with two slope changes, different from the dependence of $R_{AP}$ that shows a monotonic change. Similar behavior in $R_P$ has been reported in recent Co-based electrode MTJs with large TMR ratios [Co$_2$(Mn,Fe)Si, Co$_2$FeAl],[46,47] which contradict pure Fe electrode MTJs that show the slight monotonic reduction in $R_P$ at LT,[11,48,49,30,37] implying a difference in electronic structures between Fe and Co at MgO interfaces.

The bias voltage dependence of the differential conductance for P ($G_P$) and AP ($G_{AP}$) at RT (300 K) and 5 K are shown in Figs. 4 (a) and (b), respectively. Figures. 4 (c) and (d) shows the bias voltage dependences of the normalized TMR by the zero-bias value and the output voltage $V_{\text{out}}$ [$\equiv |V|$ × ($R_{AP} - R_P$) / $R_{AP}$, where $V$ is the bias voltage], respectively. The differential conductance was obtained by the numerical differentiation of I–V characteristics, i.e., $dI/dV$. $G_P$ spectra are asymmetric and have clear minimum structures at −0.3, −0.7, and +0.4 V, which are pronounced at 5 K. The minimum structures in the $G_P$ spectra are larger than those in typical CoFe-based MTJs.[50–52] The minimum structure of the positive bias is much deeper than that of the negative bias; the relative change from the zero-bias value reaches −34% (−29%) at 5 K (300 K). The minimum structures at the negative bias are shallower than that at the positive bias. However, the appearance of the two minima is similar to the case of the Fe/MgO/Fe(001) MTJs with RT-TMR ratios exceeding 400%.[30] Tunneling electrons primarily sense the final state, i.e., the top- (bottom-) interface electronic structures at the negative (positive) bias voltage. Thus, the $G_P$ spectra at the negative bias may represent the electronic structure of the top-MgO/CoFe(001) interface, which resembles that of a pure Fe/MgO(001) interface.[30] By contrast, the deep minimum at the positive bias may be associated with specific electronic structures of the bottom-CoFe/MgO(001) interface. These features are attributable to the much thinner CoFe insertion at the top-side interface than that at the bottom-side interface (i.e., $d_{\text{bot-CoFe}}$ = 16 ML >> $d_{\text{top-CoFe}}$ = 4 ML), implying that further improvement in the RT-TMR ratio can be expected if symmetric spectra are obtained by creating defect-free and well-balanced electronic states between the top- and bottom-side interfaces.

The bias voltage where TMR ratio reduces to half of the zero-bias value ($V_{\text{half}}$) at 300 K for the positive (negative) bias was 0.51 V (−0.49 V). Similar to common MTJs, the $V_{\text{half}}$ values at 5 K reduced to 0.28 V (−0.30 V). The curves appear nearly symmetric because of the symmetric $G_{\text{AP}}$ feature. The lower $V_{\text{half}}$ compared with Fe/MgO/Fe [10] is mainly due to the CoFe band structure; i.e., the effect of the lowered minority $\Delta_1$ band edge by Co doping into Fe, as seen in bcc-Co/MgO/bcc-Co.[45] $V_{\text{out}}$ versus bias voltage shown in Fig. 4 (d) is slightly asymmetric with respect to the bias polarity. $V_{\text{out}}$ reached 0.68 V at 300 K (0.77 V at 5 K) at the positive bias region. The value at 300 K was much larger than the values reported in CoFeB/MgO/CoFeB MTJs: 0.38 V [53] and 0.56 V.[54] In our $V_{\text{out}}$ definition, $V_{\text{out}}$ at low bias regions nearly follows the line that assumes an infinite TMR (dashed-dotted line) due to the observed large TMR ratio.

**B. MgO thickness dependence**

Figures 5 (a) and (b), respectively, show the TMR ratio and $RA$ in the P and AP states at RT for the MTJs with $d_{bot\text{-}CoFe}$ = 16 ML, $d_{top\text{-}CoFe}$ = 4 ML, and $d_{Mg}$ = 0.6 nm as a function of $d_{MgO}$ are plotted. The maximum TMR of 631% is obtained from a wafer [Fig. 2 (c)]. To obtain consistent plots, we measured the MTJ series along the $x$-direction of the wafer (MgO wedge direction) at the same $y$ position [Fig. 1 (d)]. The measurements were performed with $|V|$ < 10 mV. From Fig. 5 (a), the TMR ratio increased rapidly for $d_{MgO}$ > 1.2 nm and showed significant oscillation with $d_{MgO}$. The oscillation period was approximately 0.32 nm, which was almost identical to the values in Fe/MgO/Fe.[30,36] The maximum PV difference of 141% is significantly larger than previous reports.[10,31,36] A suppression of the oscillation at the high $d_{MgO}$ region is attributed to the deviation from the optimum interface condition due to the use of the constant $d_{Mg}$ = 0.6 nm for the entire area of the wafer. A 0.9-nm period oscillation, which was reported by Matsumoto *et al.*,[36] was not observed in our MTJs. Notably, the significant TMR and $RA$ reduction for $d_{MgO}$ < 1.3 nm in Fig. 5 is mainly because the MTJ resistances in the low $d_{MgO}$ region (less than a few Ω) are too small to neglect the effect of an electrode resistance (several Ω). From Supplementary Fig. S3 (a), the CIPT measurement shows a larger TMR ratio of ~350% with 70 Ω·μm² at $t_{MgO}$ ~ 1.2 nm.

From Fig. 5 (b), the plots of ln($RA$) show a linear increase for both P and AP states in a $d_{MgO}$ range of 1.4–2.8 nm. In addition, the slopes of both plots are almost identical. Figure 5 (c) is the close-up of Fig. 5 (a) to analyze the oscillation behavior. As performed in the previous reports,[33,36,37] the oscillation components of the $RA$s were extracted by the slope-correction of the $\ln(RA)$ plots. Here, $RA_\sigma$ was fitted by $\exp(\alpha_\sigma d_{MgO} + \beta_\sigma)$, where $\alpha_\sigma$ and $\beta_\sigma$ are the fit parameters and $\sigma$ = AP or P. We obtain the slopes $\alpha_P$ = 6.159 nm$^{-1}$ and $\alpha_{AP}$ = 6.145 nm$^{-1}$ by fits. Figures 5 (d) and (e) show the plots of the extracted oscillation components, $RA_\sigma / \exp(\alpha_\sigma d_{MgO} + \beta_\sigma)$, for the AP and P states, respectively. Both plots show significant oscillatory behavior with $d_{MgO}$ similar to the case of the TMR ratio.

The oscillation components in the TMR ratio, $RA_P$ and $RA_{AP}$, were analyzed by assuming a sinusoidal function. To determine the oscillation period, amplitude, and peak position difference for each plot, we fitted the plots using the following functions after subtracting a linear background that overlays each oscillation. The $d_{MgO}$ range to be fitted was limited to 1.6–2.1 nm to obtain consistent fitting results. The fitting functions are as follows:

$$RA_\sigma / [\exp(\alpha_\sigma d_{MgO})] = a_\sigma \sin[(2\pi / \omega_\sigma)d_{MgO} + \pi / 180° \cdot \Phi_\sigma] + (linear\ background), \quad (1)$$

$$TMR\ ratio = a_{TMR}\sin[(2\pi / \omega_{TMR})d_{MgO} + \pi / 180° \cdot \Phi_{TMR}] + (linear\ background), \quad (2)$$

where $a_\sigma$ ($a_{TMR}$) is the $RA$ (TMR) oscillation amplitude, $\omega_\sigma$ ($\omega_{TMR}$) is the $RA$ (TMR) oscillation period, and $\Phi_\sigma$ ($\Phi_{TMR}$) is the $RA$ (TMR) oscillation phase in degree, and $\sigma$ = AP or P. The fits are displayed as solid red curves in Figs. 5 (c)–(e). The obtained fitting parameters are listed in Table I. All three data show a similar oscillation period, $\omega$ = 0.323 − 0.326 nm. These oscillation periods are almost the same as the values of the TMR ratio of Fe/MgO/Fe(001) MTJs—0.32 and 0.317 nm—respectively, reported by Scheike et al.,[30] and Matsumoto et al.[36] Notably, the value is slightly larger than the 0.28 and 0.30 nm of CMS/MgO/CMS MTJs and CCFA/MgO/CCFA, respectively.[31,33] Nevertheless, the period seems to be almost independent on the used electrode materials by considering the variation in their experimental setups.

The amplitude of the TMR oscillation $a_{TMR}$ is well-matched within the fitted range (PV difference: $2a_{TMR}$ = 124% by the fit). From the $\Phi$ values, we can discuss the phase difference of $RA$ relative to the TMR ratio in degree ($2\pi$ = 360°); i.e., ($\Phi_{TMR} - \Phi_{AP}$) = −31° and ($\Phi_{TMR} - \Phi_{P}$) = −131°. Matsumoto et al.[36] obtained phase differences ($\Phi_{TMR} - \Phi_{AP}$) of −61° and ($\Phi_{TMR} - \Phi_{P}$) of 220° (= −140°) in Fe/MgO/Fe. ($\Phi_{TMR} - \Phi_{AP}$) is significantly larger in our case, whereas ($\Phi_{TMR} - \Phi_{P}$) is almost the same. Differences in these phase shifts may be associated with the difference in the magnitude of TMR ratios, or electrode materials. Nevertheless, the existence of the phase shifts results in TMR oscillation with a period almost identical to those of $R_P$ and $R_{AP}$.

To analyze the oscillations under higher bias voltages, we measured $I$–$V$ curves at RT in the P and AP states at different $d_{MgO}$ thicknesses. Figures 6 (a)–(c) show the two-dimensional (2D) maps of the $R_P$, $R_{AP}$, and TMR ratio versus $d_{MgO}$ (bottom axis) and bias voltage (left axis) for MTJs along $d_{MgO}$ with the same $y$ position, respectively. To extract the oscillation behavior, the data of $R_P$ and $R_{AP}$ were corrected using the exponential background functions, i.e., $\alpha_P$ = 6.159 nm$^{-1}$ and $\alpha_{AP}$ = 6.145 nm$^{-1}$. The TMR ratio was calculated from $R_P$ and $R_{AP}$ before the correction. Oscillations with $d_{MgO}$ at high bias voltage regions were observed in all plots. The oscillatory behavior was observed, even at high bias voltage regions (i.e., $|V| > 1$ V), suggesting the robustness of this phenomenon. This fact is surprising since the tunnel conductance and resulting TMR ratio are rather influenced by an application of such a high bias voltage. The 2D plot on $R_P$ in Fig. 6 (a) shows peak structures at the positive bias (blue areas), which agree with the $G_P$ feature in Fig. 4 (b).

We additionally measured data from rows at different $y$ positions on the wafer; the $d_{MgO}$ dependences at $V$ = 0.01, 0.2, 0.6, and 1.0 V (−0.2, −0.6, and −1.0 V) are shown in the upper (lower) panel of Figs. 6 (d)–(f). From Fig. 6 (d), the $V$ dependence of the oscillation amplitude for $R_P$ is weak. By contrast, the $V$ dependence of the amplitude for $R_{AP}$ and TMR ratio is much stronger; the amplitude is significantly suppressed at high $|V|$ regions. The $R_P$ plots show multipeak-like structures rather than a simple sine curve; such peak-like features in resistance plots were more clearly observed in a Fe/Mg$_4$Al-O$_x$/Fe MTJ at low bias.[37] The yellow dashed lines in Figs. 6 (d)–(f) indicate the change in the position of the peaks near $d_{MgO}$ = 1.8 nm with $V$ (i.e., phase shift). Significant position changes were observed for the $R_P$ (negative bias) and $R_{AP}$ (both positive and negative bias) cases, whereas changes in the TMR ratio were insignificant.

Figure 7 summarizes the maximum values [upper panels, (a)–(c)] and shifts $p$ of the first peaks near $d_{MgO}$ = 1.8 nm [middle panels, (d)–(f)], and position differences $\Delta$ between the first peaks near 1.8 nm and the second peaks near 2.1 nm [lower panels, (g)–(i), see also the lower panel of Fig. 6 (e)]. For the plots of $\Delta$ and $p$, differences normalized by the values at $V$ = 0.01 V were also indicated on the right axes. The $\Delta$ ($p$) corresponds to the oscillation period $\omega$ (peak shift $\Phi$) of Eqs. (1) and (2). Notably, the fitting using a single sine curve with Eqs. (1) and (2) was not employed for this analysis due to the observed deviation from a sine curve shape [see the $R_P$ plots in Fig. 6 (d)], scattered data, and difficulty of the subtraction of backgrounds at high bias voltages. Monotonic reduction of the peak maximum values with $|V|$ was observed for $R_{AP}$ and TMR ratio, whereas complicated behavior was observed for $R_P$; these features are expected from the bias dependences shown in Figs. 4 (a)–(c). From Figs. 7 (d) and (e), $p_P$ and $p_{AP}$ were bias dependent and they showed a slightly stronger change in the negative bias direction. This asymmetry may also reflect the difference in the electronic structures between bottom- and top-side interfaces. The changes of $p_P$, $p_{AP}$, and $p_{TMR}$ are small within ±1 V, i.e., 0.8%, 2%, and 1.5%, respectively. The trend of the $V$ dependence was opposite between $p_P$ and $p_{AP}$; i.e., $p_P$ increased with $|V|$, whereas $p_{AP}$ decreased as $|V|$ increased. These behaviors may result in an almost constant $p_{TMR}$ in a wide $V$ range. The peak difference $\Delta_P$ decreased with $V$; however, the change in $\Delta_P$ from −1 V to +1 V was only 5%. $\Delta_{AP}$ and $\Delta_{TMR}$ were almost independent on $V$. Therefore, $\Delta_P$, $\Delta_{AP}$, and $\Delta_{TMR}$ were within 0.32 ± 0.01 nm, which agreed with the values obtained from the fitting in Fig. 5 and other Fe/MgO/Fe MTJ reports.[30,36] Based on our analysis, bias voltage dependences of the oscillation components of the $R_P$, $R_{AP}$, and TMR ratio are unexpectedly weak. Notably, in the examined bias range of −0.2 to +0.2 V, the oscillation periods of CCFA/MgO/CCFA MTJs were almost bias independent for both the $R_P$ and $R_{AP}$.[33]

## IV. DISCUSSION

### A. Slopes in $RA$ versus $d_{MgO}$

The baseline of the TMR ratio shows saturation behavior at a high $d_{MgO}$ region (~550% near zero-bias). Such saturation behavior was observed in our previous MTJs with Fe/MgO/Fe [30] and Fe/Mg$_4$Al-O$_x$/Fe MTJs.[37] This behavior contradicts the coherent tunneling theories of Fe/MgO/Fe(001), where the TMR ratio monotonically increases with $d_{MgO}$ and exceeds several thousand % at several nm.[12,13,29,34,35,55,56] The TMR saturation behavior was experimentally observed in various MTJs.[10,31–33,46] The reduction in a TMR ratio at large barrier thicknesses is often observed due to localized state-assisted inelastic tunneling by the introduction of atomic defects inside the barrier.[57] However, such a process alters the magnetotransport properties, e.g., temperature dependence and $I$–$V$ characteristics, significantly depending on the barrier thickness. In the present data shown in Figs. 5 and 6, no significant change in the magnetotransport properties and oscillation behavior was observed within a wide range of $d_{MgO}$. Although a decay of the TMR oscillation amplitude was observed in CoFe/MgO/CoFe(001), the median TMR remained unchanged, indicating that dominant transport mechanisms responsible for the large TMR ratios at a low bias of our MTJ are maintained and independent of the barrier thickness, which agree with the previous Fe/MgO/Fe [30] and Fe/Mg$_4$Al-O$_x$/Fe.[37]

The TMR saturation is because the slope in ln($RA$) is almost the same in the P and AP states, i.e., $\alpha_P \sim \alpha_{AP}$ [Fig. 5 (b)]. Assuming a direct electron tunneling through a rectangular barrier within Wentzel–Kramers–Brillouin approximation (free-electron model),[58] the slope $\alpha$ is proportional to the effective barrier height.[10] Therefore, the effective barrier height of the P state is almost identical to that of the AP state. The standard model based on the $\Delta_1$-preferential coherent tunneling assumes a majority-to-majority $\Delta_1$-band tunneling process in the P state that dominates the states' conductance, resulting in a slower decay of the conductance (= 1 / resistance) in the P state with increasing barrier thickness compared with the AP state, i.e., $\alpha_P < \alpha_{AP}$. Notably, the theoretical calculations considering the interface disorder of MgO (i.e., intermixing of Mg and interfacial Fe atoms) predicts significant reduction of the TMR ratio and the appearance of the TMR saturation behavior.[59] In actual MTJs, atomic-scale imperfections near the MgO layer, such as interfacial roughness and steps, intermixing of atoms, and oxygen vacancies, cannot be perfectly excluded. Therefore, such imperfections may significantly influence TMR properties. It is expected that the introduction of barrier roughness significantly suppresses the oscillation amplitude.[39] We observed remarkable TMR oscillation with $d_{MgO}$, indicating atomically flat interfaces both at the bottom- and top-CoFe/MgO sides in a wide area (at least several tens of micrometers).

Our results suggest a MTJ structure that is closer to an ideal structure used for theoretical calculation assuming a "free-electron" barrier (Fe/barrier/Fe), in which the relationship of $\alpha_P \approx \alpha_{AP}$ is reproduced.[60] This implies that further enhancement in a TMR ratio may be achievable if the $\Delta_1$-preferential tunneling is enhanced further. Therefore, improving (001)-orientation and suppressing atomic defect formations, such as oxygen vacancies and misfit dislocations, may lead to large TMR ratios by activating the coherent tunneling mechanism.

**B. *RA* oscillation behavior**

The TMR oscillation behavior is strongly linked to the resistance oscillation components in the P and AP states. All oscillation periods are almost similar (0.31–0.33 nm) in our MTJs. Further, TMR oscillations are observed in all MTJs during the optimization of the MTJ stack, e.g., tuning of growth parameters and thicknesses of the FM and insertion layers. The oscillation amplitudes increase with increasing maximum TMR ratio, i.e., improved effective spin polarization. The amplitude is further enhanced in a MTJ with a composition-tuned tunnel barrier, $Mg_4Al-O_x$ (MAO),[37] compared with one with a MgO barrier.[30] The introduction of the MAO barrier can effectively reduce misfit dislocations at barrier interfaces,[61] resulting in an enhancement due to improvement in the barrier interface states. These facts indicate the robustness of the phenomenon.

The oscillation periods are similar to the previous experimental values for various MTJs with bcc-structured electrodes (0.30 ± 0.02 nm).[10,36] This universality of the ~0.3 nm period contradicts many theoretical calculations so far, e.g., no oscillations,[29,35] different oscillation periods,[12,39] and an oscillation period sensitive to interface roughness and atomic vacancies.[38] Our measurements showed that the oscillations in $R_P$, $R_{AP}$, and a TMR ratio persist, even at high bias voltages exceeding ±1 V, suggesting that direct effects on transport through sharp electronic states, such as interface resonant states [10,13] and quantum resonant states, in FM layers [39] may not be the main origins. To understand the behavior of the robust TMR oscillation effect, the appearance of the oscillation component and weak bias voltage dependence of its period require theoretical clarifications.

Although the oscillation behavior as a whole is highly robust, we observed a slight change with bias voltage in the resistance oscillation period and phase (peak position) between the P and AP states. In addition, the slope-corrected $R_P$ plots show peak-like structures rather than a simple sine curve. Such behaviors may be crucial to understanding the phenomenon because the resistances determine features of the TMR oscillation. In fact, a TMR ratio of Fe/Mg$_4$Al-O$_x$/Fe(001), which exhibits a more significant oscillation than Fe/MgO/Fe(001), shows sawtooth-like curves.[37] The insets in Figs. 7 (e) and (h) are plots of period differences ($p_{AP} - p_P$) and peak position differences ($\Delta_{AP} - \Delta_P$), respectively. ($\Delta_{AP} - \Delta_P$) slightly increased with $V$, but the change was insignificant. Meanwhile, ($p_{AP} - p_P$) depended more clearly on the bias voltage. More systematic experiments will help clarify this phenomenon. In addition, the effect of nanostructural reconstruction at the barrier interface requires investigation, especially when a noninteger monolayer thick barrier layer [i.e., 1 ML = 0.21 nm for MgO(001)] is used.

## V. CONCLUSIONS

We observed giant TMR ratios of up to 631% and 1143% at RT and 10 K, respectively, using a CoFe/MgO/CoFe(001) MTJ. The very large values are attributed to improved (001)-orientation and MgO barrier interface crystallinity by the introduction of ultrathin Mg and CoFe insertion layers and control of barriers' oxidation states. Further, the observed TMR ratios showed significant oscillation behavior as a function of $d_{MgO}$. The oscillation PV difference reached 141% at RT. The oscillations were observed, even at high bias voltages exceeding ±1 V, indicating that the oscillation dominates the transport properties of the MTJs. The oscillation component of $R_P$ shows peak-like shapes. Such peak-like oscillation components were also observed in the recent Fe/Mg$_4$Al-O$_x$/Fe(001). Therefore, a simple sinusoidal function is not sufficient to reproduce the signals of MTJs exhibiting giant TMR ratios. In addition, our MTJ shows TMR saturation behavior at high $d_{MgO}$; namely, the slope of $\ln(R_{AP})$ with $d_{MgO}$ is almost identical to that of $\ln(R_P)$, which contradicts theoretical calculations that generally show a monotonic increase in TMR ratio with $d_{MgO}$ due to an enhanced $\Delta_1$ filtering effect. Therefore, TMR oscillation and saturation in our MTJs cannot be explained only by a common $\Delta_1$ band preferential tunneling effect. Our results will be relevant to further investigate TMR physics in depth toward the true mechanism.

Although we observed a large TMR ratio and its oscillation, asymmetric $I$–$V$ characteristics (i.e., bias voltage dependences of $dI/dV$ and TMR ratio) were still observed, indicating further improvement in RT-TMR ratios is expected if more symmetric $I$–$V$ characteristics are achieved by tuning barrier interfaces, such as interface crystallinity improvement by nanoinsertions and use of lattice-matched barrier, e.g., MgAl$_2$O$_4$. Because electrical signal outputs and signal separation between low and high resistance states improve with TMR ratio, the achievement of giant RT-TMR ratios will significantly expand the possibility of future device generation, such as large-capacity MRAMs (i.e., multivalue MRAMs), ultrahighly sensitive magnetoresistance sensors, and MTJ-based artificial neural networks. Our demonstration of the giant RT-TMR ratios in a wide range of $RA$ will be an essential step for developing such MTJ-based applications.

ACKNOWLEDGEMENTS

The authors are grateful to Shinji Yuasa for his valuable comments on TMR oscillation effect of MgO-based MTJs. The authors thank to Yoshio Miura and Keisuke Masuda for their fruitful discussion from theoretical viewpoints, and to Hiromi Ikeda for her technical support on device microfabrication. The work was partly supported by the ImPACT Program of the Council for Science, Technology and innovation (Cabinet Office, Government of Japan), JSPS KAKENHI Grant Nos. 16H06332, 21H01750, and 21H01397. This paper is partly based on results obtained from a project, JPNP16007, commissioned by the New Energy and Industrial Technology Development Organization (NEDO).

**Table**

**Table I.** Fitting results with ±1 standard deviation obtained from the data of Fig. 5, where $a$ is the amplitude, $\omega$ is the period, $\Phi$ is the phase shift in degree, $\alpha$ is the $RA$ slope, and $\beta$ is the $RA$ intercept.

|  | $a$ | $\omega$ (nm) | $\Phi$ (°) | $\alpha$ (nm$^{-1}$) | $\beta$ |
|---|---|---|---|---|---|
| TMR | 61.97±2.09 (%) | 0.325±0.004 | 192±22 | – | – |
| $RA_{AP}$ | 0.0136±0.0015 | 0.323±0.010 | 223±66 | 6.145 | −5.310 |
| $RA_P$ | 0.0019±0.0001 | 0.326±0.009 | 323±58 | 6.159 | −7.197 |

**Figures**

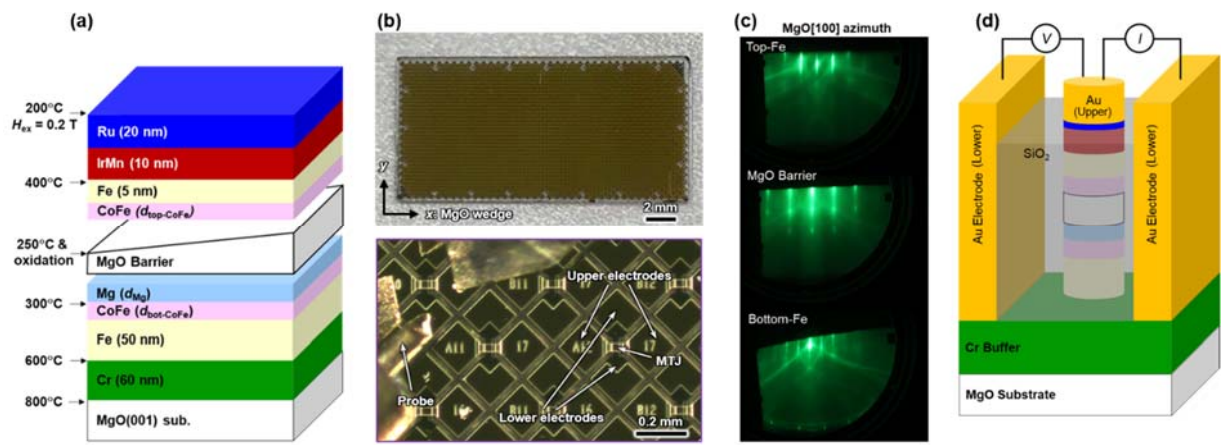

**FIG. 1.** (a) Schematic MTJ stacking structure and processes. (b) Images of a MTJ wafer. (c) RHEED patterns along MgO[100] azimuth of top-Fe (upper), MgO barrier (middle), and bottom-Fe (lower). (d) Schematic structure of a patterned MTJ pillar after microfabrication.

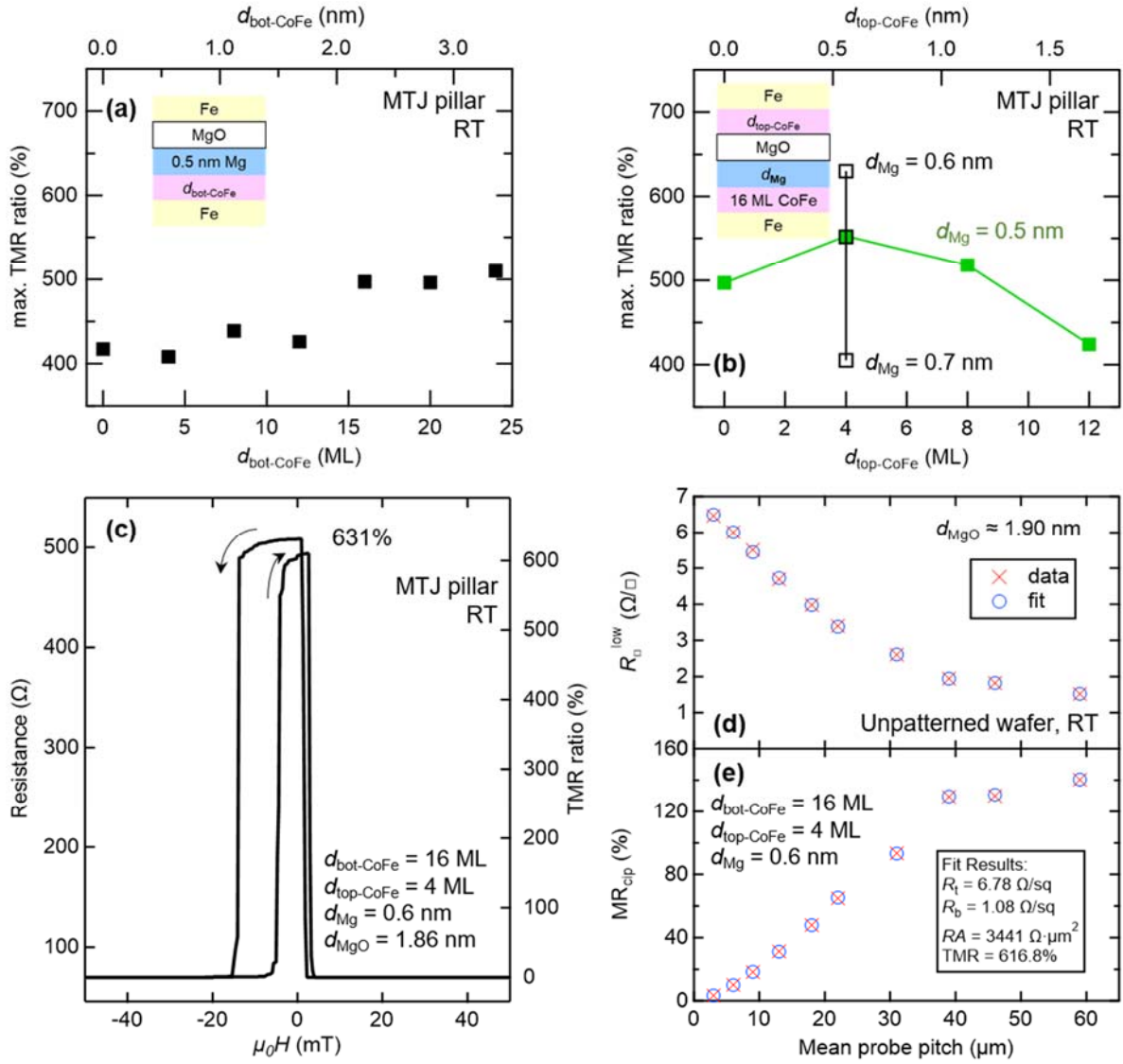

**FIG. 2.** (a) Maximum RT-TMR ratio versus bottom-CoFe insertion thickness $d_{\text{bot-CoFe}}$. (b) Maximum TMR ratio versus top-CoFe insertion thickness $d_{\text{top-CoFe}}$ and Mg insertion thicknesses $d_{\text{Mg}}$. Insets of (a) and (b) show schematic stacking structures. (c) Resistance (left axis) and TMR ratio (right axis) versus magnetic field $\mu_0 H$ of a MTJ with the maximum RT-TMR ratio ($d_{\text{bot-CoFe}}$ = 16 ML, $d_{\text{top-CoFe}}$ = 4 ML, $d_{\text{Mg}}$ = 0.6 nm, and $d_{\text{MgO}}$ = 1.86 nm). (d) CIPT results of the unpatterned wafer of (c): $R_\square^{\text{low}}$ (upper) and MR$_{\text{cip}}$ (lower). Other CIPT results of the wafer are shown in Supplementary Figs. S2 and S3.

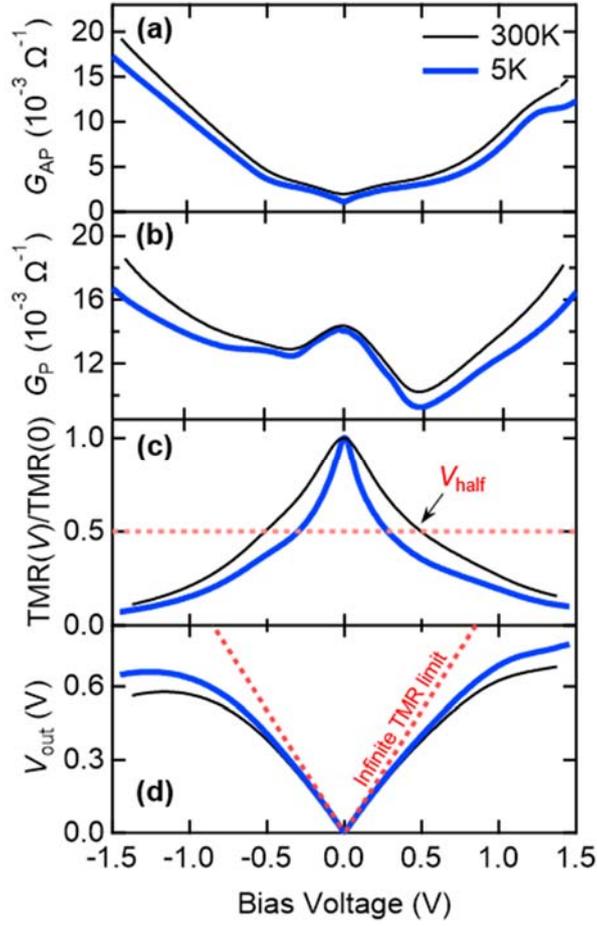

**FIG. 3.** Temperature dependences of TMR properties of the MTJ with $d_{\text{bot-CoFe}}$ = 16 ML, $d_{\text{top-CoFe}}$ = 4 ML, $d_{\text{Mg}}$ = 0.6 nm, and $d_{\text{MgO}}$ = 1.86 nm. (a) TMR ratio (left axis) and conductance ratio (right axis). (b) $R_{\text{P}}$ (left axis) and $R_{\text{AP}}$ (right axis). Inset shows TMR ratio versus magnetic field $\mu_0 H$ at 10 K.

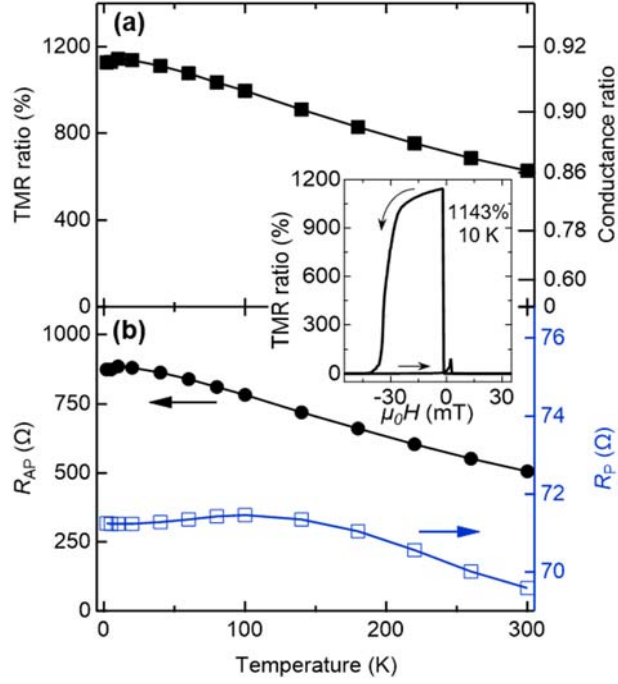

**FIG. 4.** Bias voltage dependence of a MTJ with $d_{\text{bot-CoFe}}$ = 16 ML, $d_{\text{top-CoFe}}$ = 4 ML, $d_{\text{Mg}}$ = 0.6 nm, and $d_{\text{MgO}}$ = 1.86 nm measured at 300 K (RT) and 5 K. (a) and (b) Differential conductance $dI/dV$ of AP and P states, respectively. (c) TMR ratio normalized by zero-bias value. (d) Output voltage $V_{\text{out}} \equiv |V| \times (R_{\text{AP}} - R_{\text{P}}) / R_{\text{AP}}$.

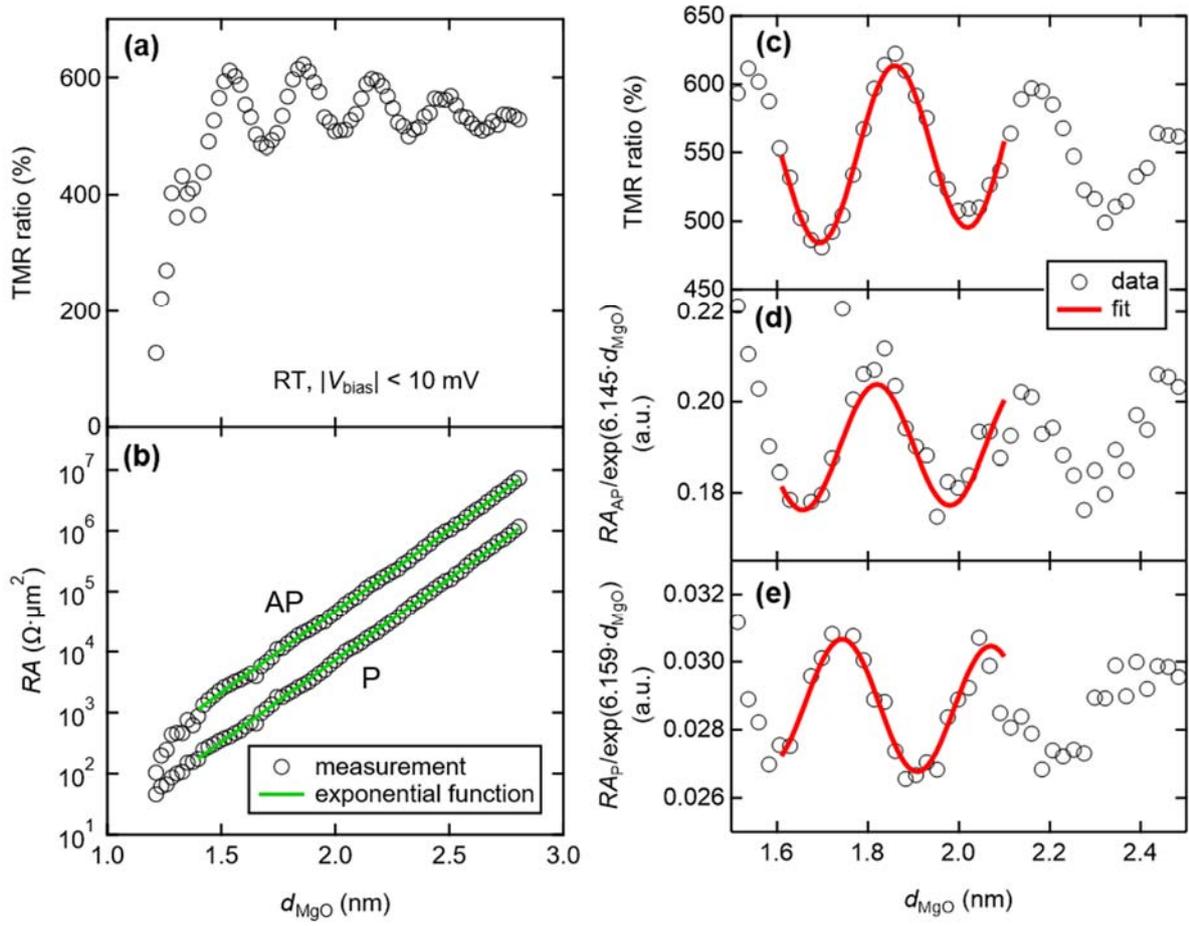

**FIG. 5.** $d_{MgO}$ dependences of (a) TMR ratio and (b) $RA$ for P and AP states at RT. Green lines are fits using an exponential function. (c) Close-up of the plot of TMR ratio. (d) and (e) Background corrected $RA$ plots for AP and P states, respectively. Red lines are fits using Eqs. (1) and (2) in $d_{MgO}$ range of 1.6–2.1 nm.

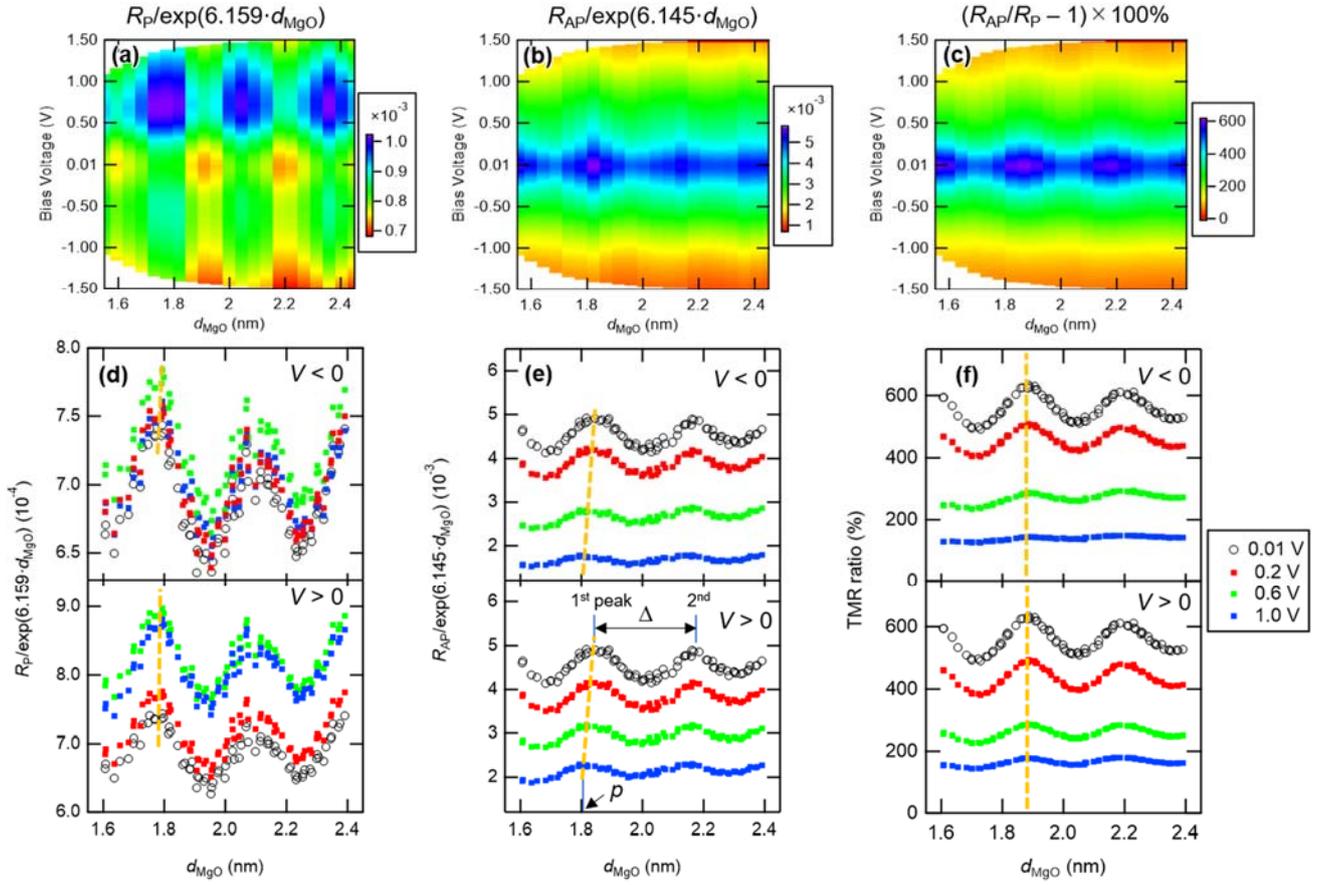

**FIG. 6.** (a)–(c) 2D maps of the exponential background corrected resistance $R_P$, $R_{AP}$, and TMR ratios, respectively, as a function of bias voltage (left axis) and $d_{MgO}$ (bottom axis). (d)–(f) Respective slices from the 2D maps with additional data points at fixed bias voltages. Dashed yellow lines are a guide for the eye showing the shift of the first peak position, i.e., phase shift. Annotations represent the values obtained for Fig. 7.

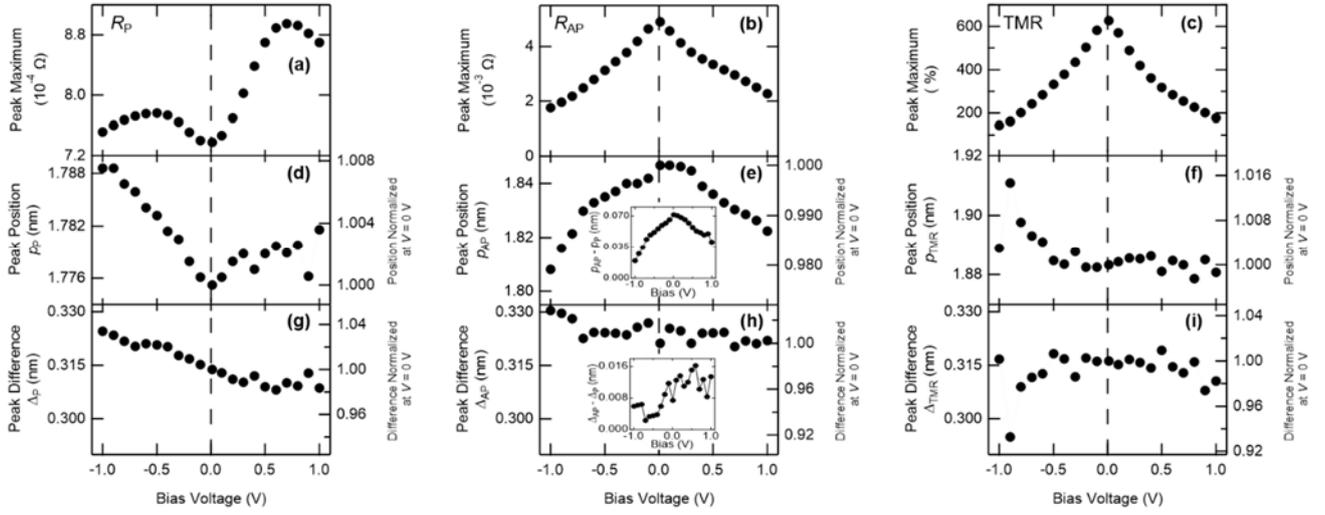

**FIG. 7.** Values of (a)–(c) the first peak maximum, (d)–(f) first peak position $p$, and difference between the second and (d)–(f) first peak position $\Delta$ for $R_P$, $R_{AP}$, and TMR ratios obtained from Figs. 6 (g)–(i), respectively. Insets in (e) and (h) show the respective differences between the AP and P values.

# Supplementary Material

**S1. Other examples of current in-plane tunneling (CIPT) measurements**

Before magnetic tunnel junctions (MTJs) were patterned into micrometer-scale pillars via microfabrication processes, CIPT measurements were performed for unpatterned MTJ wafers to exclude any possible measurement or microfabrication errors as in our previous Fe/MgO/Fe(001) study.[1] We used a wide-type M12PP Micro Twelve-Point Probes (Capres A/S, M12PP_005) to obtain reliable fit calculations for a resistance area ($RA$) range of ~100–20,000 $\Omega \cdot \mu m^2$ ($d_{MgO}$ range: 1.2–2.2 nm, see the Supplementary Material of Ref.[1]). The probe has a minimum (maximum) mean probe spacing of 3.0 (59.0) μm. By choosing four of twelve microprobes, ten different probe spacings can be selected. The sheet resistance $R_\square$ and CIP magnetoresistance ratio ($MR_{CIP}$) for each spacing are measured; then, the tunnel magnetoresistance (TMR) ratio, $RA$, bottom electrode's sheet resistance ($R_b$), and top electrode's sheet resistance ($R_t$) at zero-bias voltage of an unpatterned MTJ were obtained from the best fits of the theoretical function.[2]

The CIPT result of a MTJ wafer (CIPT-TMR ratio: 616.8%) with $d_{bot\text{-}CoFe}$ = 16 ML, $d_{top\text{-}CoFe}$ = 4 ML, and $d_{Mg}$ = 0.6 nm is shown in Figs. 2 (d) and (e). Figure S1 shows CIPT results of other MTJs with $d_{Mg}$ = 0.5 nm and different CoFe thicknesses as examples: $d_{bot\text{-}CoFe}$, $d_{top\text{-}CoFe}$ = (a) 16, 0 ML, (b) 24, 0 ML, and (c) 16 ML, 4 ML. All experimental data are well-fitted by theoretical curves, and obtained $R_t$ and $R_b$ agree with the design electrode stacks. Notably, sample (c) has a thinner Ru cap layer, yielding a larger $R_t$ value of 23 $\Omega/\square$. TMR ratios >400% for samples (a) and (b) and >500% for sample (c) were obtained. The CIPT-TMR ratios are slightly smaller than that of microfabricated ones. This underestimation is likely because a perfect AP state is not well-obtained in the unpatterned wafers, as the CoFe insertion (i.e., CoFe/Fe/IrMn) reduces the exchange bias field of the top layer and changes the magnetic easy axis from the magnetic field direction ∥ Fe[100] (i.e., easy axis: Fe[100], $Co_{50}Fe_{50}$[110]).

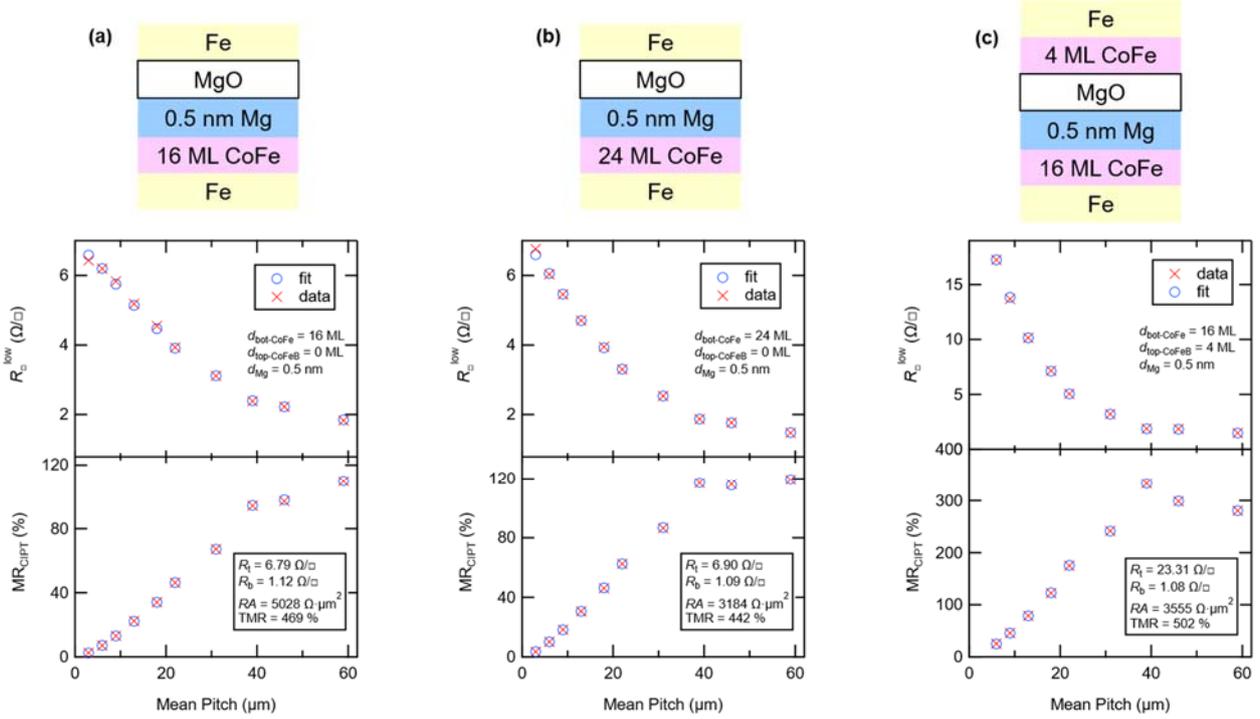

**FIG. S1.** Examples of CIPT measurements (red cross marks) and fit results (blue open circles) at room temperature of MTJs with different CoFe insertion conditions: (a) $d_{\text{bot-CoFe}}$ = 16 monolayer (ML), $d_{\text{top-CoFe}}$ = 0 ML, (b) $d_{\text{bot-CoFe}}$ = 24 ML, $d_{\text{top-CoFe}}$ = 0 ML, and (c) $d_{\text{bot-CoFe}}$ = 16 ML and $d_{\text{top-CoFe}}$ = 4 ML, as shown in schematics of their stack structures. $d_{\text{MgO}}$ is approximately 1.9 nm. Upper graphs are sheet resistances per square in parallel state ($R_\square^{\text{low}}$), and lower graphs are CIPT magnetoresistance ratios (MR$_{\text{CIPT}}$) with fits by the theoretical model.[2]

## S2. CIPT results of MgO thickness $d_{\text{MgO}}$ dependence of CoFe/MgO/CoFe MTJs

In Fig. 5 (a), the CoFe/MgO/CoFe MTJ shows a significant reduction in the TMR ratio at a low $d_{\text{MgO}}$ region (<1.3 nm). As noted in the main text, low device resistances by low $RA$ values < 100 Ω·μm$^2$ and the 10 × 5 μm$^2$ MTJ pillar size (~39 μm$^2$) affect the evaluated TMR ratio and $RA$ values. Our MTJ device structure shown in Fig. 1 (c) has an electrode resistance of several Ω, making it difficult to obtain accurate $RA$ values for MTJs with a low resistance of less than a few Ω. To evaluate more reliable TMR ratios at the low $RA$ range, CIPT measurement results were compared with the DC 4-probe measurement results of Fig. 5. Figures S2 (a) and (b) show the TMR ratio and $RA$ as a function of $d_{\text{MgO}}$, respectively. For CIPT data plots, 0.1 nm was added to the original $d_{\text{MgO}}$ to compensate for slight wafer rotation during the wedged MgO deposition (i.e., small misalignment between the substrate and linear shutter) and our CIPT scan position accuracy. The CIPT-TMR ratio basically follows the 4-probe data in the $d_{\text{MgO}}$ range of 1.3–2.2 nm. Fit parameters, $R_t$ and $R_b$, are nearly constant for this $d_{\text{MgO}}$ range, indicating that the CIPT measurement data were well-fitted by the model. A CIPT-TMR ratio at $t_{MgO} \sim 1.2$ nm shows ~350% with 70 Ω·μm², indicating that the actual TMR ratio without the electrode resistance effect can be larger than observed in the 4-probe measurement at the low $t_{MgO}$ region. Notably, as mentioned in Sec. S1, some CIPT-TMR ratios of the CoFe/MgO/CoFe are smaller than the 4-probe TMR ratios due to the instability of the AP state before the MTJ patterning. Therefore, some of our CIPT-TMR ratios were underestimated. Nevertheless, the TMR oscillation behavior is observed in the CIPT-TMR ratio plot, similar to the Fe/MgO/Fe case.[1] Fig. S2 shows the scan data of a wafer Y direction (perpendicular to the MgO wedge) at $t_{MgO} \sim 1.9$ nm with hexagon marks. Although the variation in the TMR ratio and $RA$ was observed due to the substrate–shutter misalignment, giant TMR ratios >600% were observed from some measured positions. Figures S3 shows two examples of CIPT fit results obtained from positions at $t_{MgO}$ of (a) ~ 1.2 nm and (b) ~ 1.9 nm.

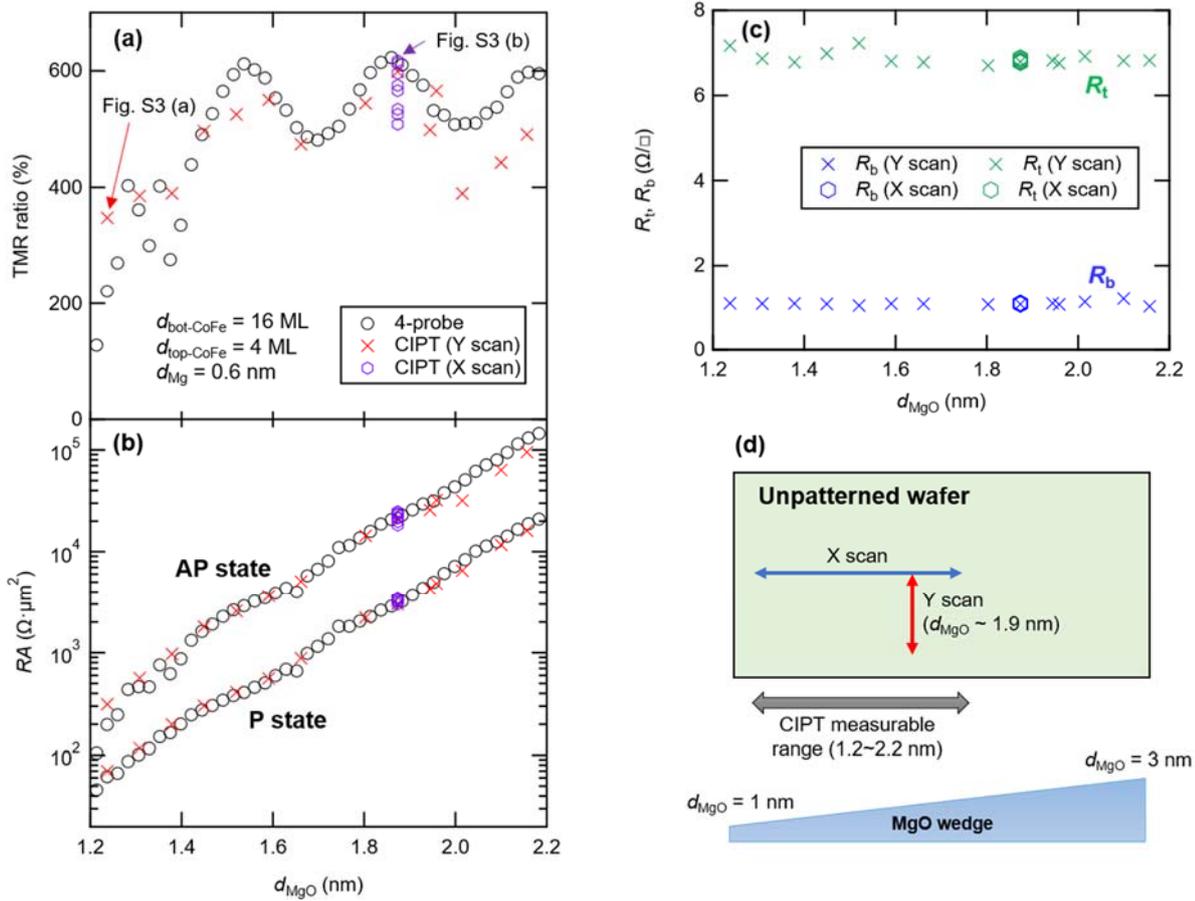

**FIG. S2.** (a) and (b) Comparison between CIPT measurements (red cross marks and purple hexagons) and DC 4-probe measurements of patterned MTJs (black open circles, replotted using Fig. 5) at RT for $d_{bot\text{-}CoFe}$ = 16 ML, $d_{top\text{-}CoFe}$ = 4 ML, and $d_{Mg}$ = 0.6 nm: (a) TMR ratio, (b) $RA_P$ and $RA_{AP}$. (c) $R_t$ and $R_b$ values of the CIPT fits. (d) Schematic of the wafer and scan directions for CIPT measurements. Red cross marks (purple hexagons) of (a) and (b) are X (Y) scan data.

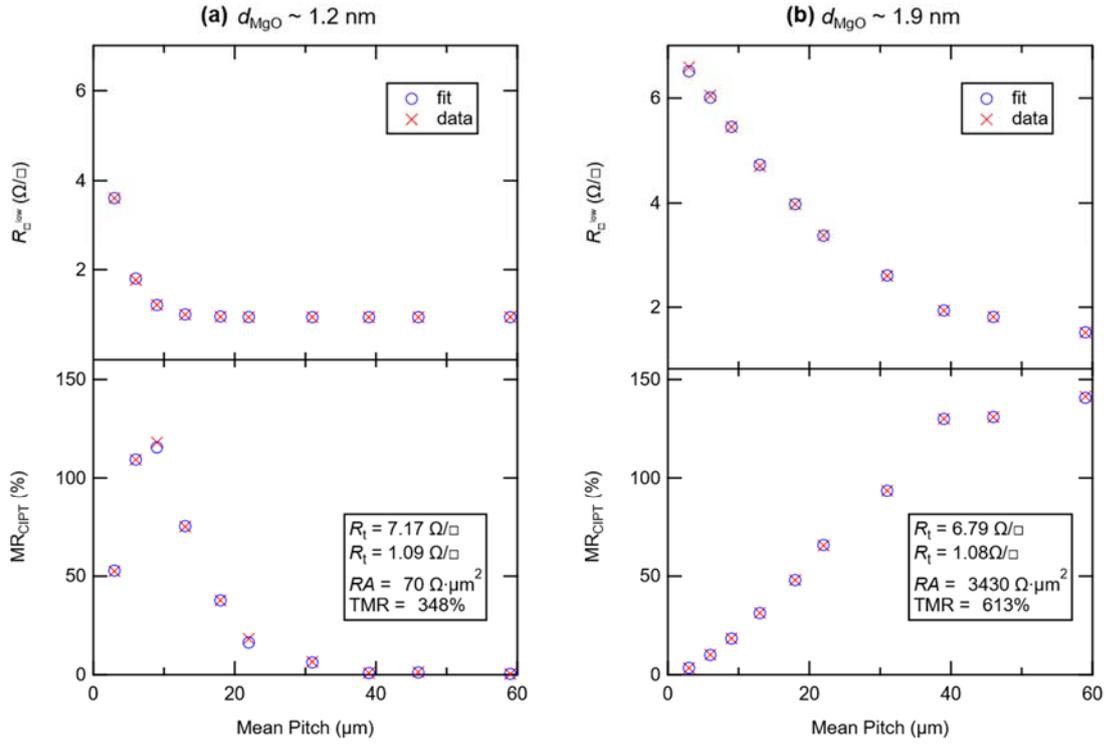

**FIG. S3.** CIPT fit results of two MTJs of Fig. S2: (a) $d_{MgO}$ ~ 1.2 nm and (b) ~ 1.9 nm, different Y position of Figs. 2 (d) and (e).